\pdfoutput=1
\documentclass[aps,twocolumn,prb,showpacs,floatfix]{revtex4}
\usepackage{graphicx}
\usepackage{amsmath,amsfonts,amssymb,bm}
\usepackage{times}
\usepackage{dcolumn}
\begin{document}

\title{Trapping of electrons near chemisorbed hydrogen on graphene}
\author{J. A. Verg\'es}
\email{jav@icmm.csic.es}
\author{P.L. de Andres}
\affiliation{
Instituto de Ciencia de Materiales de Madrid (CSIC), Cantoblanco, 28049 Madrid,
Spain.}
\date{22 May 2009}
\begin{abstract}
Chemical adsorption of atomic hydrogen on a negatively charged single 
layer graphene sheet has been analyzed with {\it ab-initio}
Density Functional Theory calculations. 
We have simulated both finite clusters and infinite periodic systems 
to investigate the effect of different ingredients of the theory, 
e.g. exchange and correlation potentials, basis sets, etc.
Hydrogen's electron affinity dominates the energetic balance in the charged 
systems and the extra electron is predominantly attracted to a region 
nearby the chemisorbed atom. The main consequences are: 
(i) the cancellation of the unpaired spin resulting in a singlet 
ground-state, 
and (ii) a stronger interaction between hydrogen and the graphene sheet. 
\end{abstract}
\pacs{73.20.Hb,68.43.Bc,82.65.+r,75.70.Rf}
\keywords{
graphene, hydrogen, chemisorption,
complex defect center,
electronegativity of hydrogen on graphene,
charged hydrogen groundstate on graphene}
\maketitle

\section{Introduction}
Planar two-dimensional graphene has been considered to be a very promising 
new material since its preparation by
Novoselov {\em et al.}\cite{Novoselov} and Berger {\em et al.}\cite{Berger}
in 2004. Recent interest is focused on the appearance of magnetism
around point defects in graphene\cite{magnetism} and
the possibility of hydrogen storage\cite{HonGraphene,stress,HonGraphene2,
graphane}. A global understanding of the origin of magnetism in finite
graphene systems is provided by Lieb's theorem on
bipartite lattices\cite{lieb}.
Any unbalance between the numbers of sites belonging to each of the two
sublattices gives rise to a magnetic groundstate\cite{innombrables}.
This result rests on the validity of a simple Hubbard Hamiltonian which
certainly works for the semi-quantitative description of the spin states
of some polycyclic aromatic hydrocarbon molecules but not for the
corresponding charged states\cite{PAH}.  
The saturation of a carbon $\pi$-electron by hydrogen is one of the simplest
ways to change the balance between sublattice sites and, consequently,
it produces spin polarization in the neighborhood. This provides an
interesting link between hydrogenation and spin production that
motivates this work. 
We shall demonstrate that the 1/2 spin originated by the
presence of an isolated hydrogen atom on top of a carbon atom
belonging to planar graphene can be
quenched if an extra electron bounds to the defect. Numerical
results obtained by {\em ab initio} open-shell
Density Functional Methods (DFT) suggest that the complex defect
is energetically favorable. A similar spin quenching phenomenon
was discussed some years ago by
Duplock, Scheffler, and Lindan for hydrogen near
a Stone-Wales defect\cite{SW}. These authors have shown that the
spin polarized groundstate around chemisorbed hydrogen disappears
in the presence of a Stone-Wales defect. In their interpretation,
this result is probably due to the strong destruction of alternation
near the defect that eliminates the tendency to antiferromagnetic order.
In our case, however, we argue that the mere flow of charge is enough
to heal the local unbalance between sublattices due to the
existence of one chemisorbed hydrogen on an otherwise ideal
graphene material.

The rest of the paper is organized as follows: The {\it ab initio} methods
used in this work are presented in section II
followed by the discussion of our main numerical results.
Final section III just remarks our main message.

\begin{table*}
\caption{
Total energies (in hartree) for atomic hydrogen and its anion,
for coronene $({\rm C}_{24}{\rm H}_{12})$ and coronene anion
$({\rm C}_{24}{\rm H}_{12})^{-}$
both ideal and with a chemisorbed H atom and for a larger cluster
$({\rm C}_{54}{\rm H}_{18})$ sometimes called supercoronene
and its anion both planar and deformed by the presence of a chemisorbed
H atom. Since results are given for two or more gaussian basis sets
some rough estimation of error is possible.} 
\begin{ruledtabular}
\begin{tabular}{lccccc}
Cluster & Total Spin & Energy(MIDI) & Energy(CCT) & Energy(PC2) & Energy(PC3) \\
\hline
${\rm H}$       & $\frac{1}{2}$ 
& -0.4953 & -0.4988 & -0.4990 & -0.4991 \\
$\rm H^{-}$   & 0 
& -0.4602 & -0.5035 & -0.5177 & -0.5254 \\
${{\rm C}_{24}{\rm H}_{12}}$   & 0 
& -915.9342 & -921.6070 & -921.6516 & -921.6950 \\
$({\rm C}_{24}{\rm H}_{12})^{-}$ & $\frac{1}{2}$ 
& -915.9342 & -921.6186 & -921.6660 & -921.7097\\
${{\rm C}_{24}{\rm H}_{13}}$ & $\frac{1}{2}$ 
& -916.4438 & -922.1269 & -922.1723 & - \\
$({\rm C}_{24}{\rm H}_{13})^{-}$ & 0 
& -916.4853 & -922.1812 & -922.2300 & - \\
${{\rm C}_{54}{\rm H}_{18}}$   & 0 
& -2055.5931 & - & -2068.3971 & - \\
$({\rm C}_{54}{\rm H}_{18})^{-}$ & $\frac{1}{2}$ 
& -2055.6323 & - & -2068.4469 & - \\
${{\rm C}_{54}{\rm H}_{19}}$ & $\frac{1}{2}$ 
& -2056.1072 & - & -2068.9095 & - \\
$({\rm C}_{54}{\rm H}_{19})^{-}$ & 0 
& -2056.1868 & - & -2068.9919 & - \\
\end{tabular}
\end{ruledtabular}
\label{TotalEnergies}
\end{table*}

\section{{\it Ab initio} calculations: Methods and Results}
\begin{figure}
\includegraphics[width=\columnwidth]{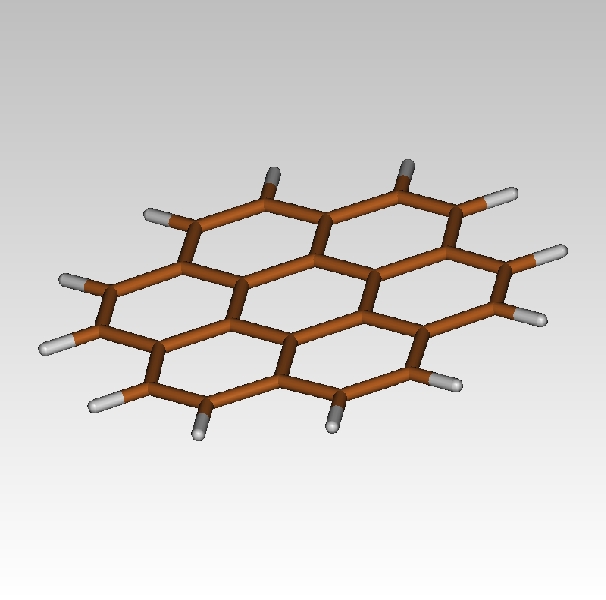}
\includegraphics[width=\columnwidth]{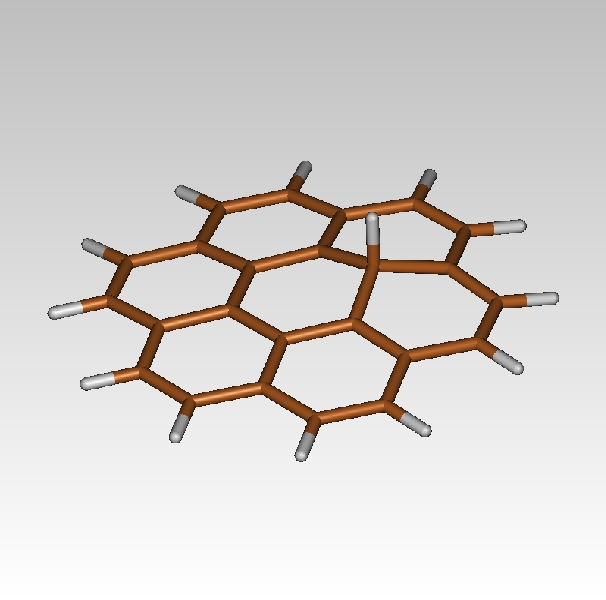}
\caption{(color online)
Upper panel: schematic representation of a coronene molecule showing
a honeycomb lattice inner structure saturated in the boundary
by hydrogen atoms so the coordination of carbon atoms
is preserved over the whole system.
Lower panel: An extra hydrogen atom is chemisorbed
on top of a carbon atom belonging to the inner ring of coronene.
Although the sp$^2$ to sp$^3$ reconstruction is only faintly visible
in the figure, the C-H bonding distance and the
details of the upward relaxation of C neighboring atoms
coincide with those
given in Ref.(\onlinecite{stress}).}
\label{Coronene}
\end{figure}

\begin{figure}
\includegraphics[width=\columnwidth]{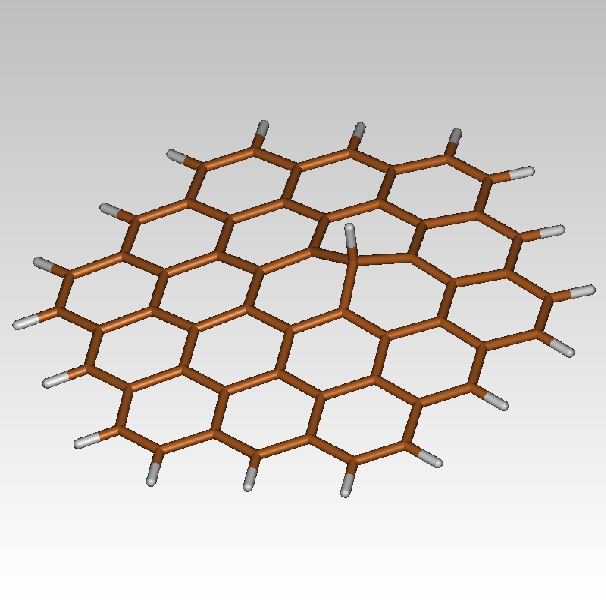}
\caption{(color online)
Hydrogen chemisorbed on supercoronene (${{\rm C}_{54}{\rm H}_{18}}$)
}
\label{Htop-SC}
\end{figure}

Two planar carbon clusters with the structure of graphene have been
chosen to calculate the energetics of hydrogen absorption both at the
neutral state and when the system is electronically charged 
with an extra electron (anions).
The first system is the well known coronene polycyclic aromatic
hydrocarbon (PAH) represented in Fig.~\ref{Coronene}
while the second one is a larger PAH obtained from coronene adding an
extra shell of benzene rings. 
This system is referred as supercoronene in the
literature. We notice that 
this larger PAH molecule has not yet been synthesized 
but nonetheless it provides a good theoretical benchmark for our purposes
(Fig.~\ref{Htop-SC} schematically shows one hydrogen 
chemisorbed on supercoronene).
The groundstate for both PAH's does not show spin polarization
(total spin is zero).
This is an important difference with our previous study of hydrogen
chemisorption on graphene where spin one-half clusters were used to
preserve the point symmetry and facilitate the
computational effort\cite{stress}.
In the present work, however, we focus on the description
of spin polarization and we must start with an unpolarized
cluster to accurately simulate the graphene layer. Point symmetry is lost and
computational load is larger. Nevertheless, we have checked that
both structural and energetic results for chemisorbed H coincide
with the values given in our previous work\cite{stress}.

Quantum chemistry calculations have been done using the GAMESS
program \cite{GAMESS}. Several sets of gaussian basis functions
have been employed. Depending on the computational effort
we have been able to assess the convergence of
numerical results for some cases by comparing results obtained with bases of
different sizes. For the largest systems, however,
we have been forced to choose a minimal basis and convergence
could only be assessed by reference to the smaller clusters.
Specifically,
we have always started by trying the so-called MIDI basis\cite{MIDI},
and when possible we have moved to a correlation consistent basis 
referred in the literature as cc-pVTZ\cite{CCT}
(CCT within GAMESS and this paper) and a DFT adapted hierarchy of basis
called PCn where n indicates the level of polarization\cite{PCn}.
Our best results correspond to the larger PCn basis that we have
been able to use in each case.
Unrestricted Hartree Fock (UHF) calculations
have been intentionally avoided because total spin of the wavefunction
is undefined in those cases. Therefore, HF calculations for an odd
total number of electrons have been performed using the Restricted-Open-Shell
variant. All results presented in this work for clusters have been obtained
using the Becke-Lee-Yang-Parr hybrid density functional RB3LYP \cite{B3LYP}.

Our choice of finite clusters of various sizes poses the question of
to what extent results are affected by the particular boundary conditions
imposed to solve the quantum problem for electrons. Therefore, we check
by comparing with calculations performed on extended
periodic models using periodic boundary conditions 
and a plane-waves basis. This model is set up
so a single H atom is adsorbed on a $4 \times 4$ graphene supercell 
including 32 carbon atoms on a honeycomb lattice 
(see Fig.~\ref{fig4x4},
$a=b=9.84$ {\AA}, $c=23.4$ {\AA}, $\alpha=\beta=90^{o}$,
$\gamma=60^{o}$). 
We use ultrasoft pseudopotentials\cite{Vanderbilt},  
an energy cutoff of 310 eV, 
and a Monkhrost-Pack mesh of $3 \times 3 \times 1$\cite{monk}. 
Actual calculations are performed with the 
CASTEP program allowing for spin polarization of
the different electronic bands\cite{Payne,Accelrys}. 
To describe the exchange and correlation potential
we use the local density 
approximation\cite{Kohn}.
Forces and stresses on the system are converged to the
usual thresholds ($9\times 10^{-3}$ eV/{\AA} and $0.01$ GPa)
and the total energy is minimized for the different
systems. This procedure cannot provide an accurate
description for the electron affinity energy since the extra
electron in the supercell is intentionally neutralized with
a uniform positive background to subtract infinite contributions
in the periodic system. This uniform background, however, has
little effect on the spatial distribution of the electronic
and spin densities, that
can be analyzed with reasonable confidence. 

Table I compiles the bulk of our quantum-chemistry results. Notice
that together with the total energies needed to get chemisorption
energies for hydrogen on graphene, we have computed the energies
corresponding to systems charged with an extra electron (anions in the
molecular case). Total spin of the groundstate is given by the second
column of the table. It can be seen that the S=0 value of the neutral
molecules is recovered by the anions of the hydrogenated cases. 
Table II gives the electron affinities that are obtained from
the results shown in Table I\cite{EA}. Nice convergence to the experimental
electron affinity of hydrogen is observed in the first entry of the table. 
Results for larger clusters are limited by our computational means 
but nonetheless our results at the PC2 level
are good enough to support our conclusions.
A word of caution is in order here: although these results for the spin seem 
to fit nicely within a simple electron-counting scheme (i.e., zero spin for 
even number of electrons and net spin for odd number of electrons) this is not 
always true. In particular, we recall the case where two hydrogen atoms are 
adsorbed on the graphene layer: while adsorption of the two hydrogens in 
next-neighbor positions results in a ground state with no net spin, 
adsorption in next to next-neighbor sites results in a ground state with 
a net $S=2$ $\mu_{B}$. 
This is related to the fact that graphene is a bipartite lattice, 
and it is in accordance to Lieb's theorem\cite{lieb}, showing the limitations
of simple electron-counting rules.

\begin{table}
\caption{
Electron affinities of hydrogen, coronene, monohydrogenated coronene,
supercoronene and monohydrogenated supercoronene
obtained from the results compiled in Table I. 
Energies are now given in eV.}
\begin{ruledtabular}
\begin{tabular}{lcccc}
Cluster & Energy(CCT) & Energy(PC2) & Energy(PC3) & \mbox{Experimental} \\
\hline
$\rm H$   &  0.13  &  0.51  &  0.72  &  0.75\footnotemark[1] \\
${{\rm C}_{24}{\rm H}_{12}}$ & 0.32 & 0.39 & 0.40 & 0.47\footnotemark[2] \\
${{\rm C}_{24}{\rm H}_{13}}$ & 1.48 & 1.57 &   -  &   -  \\
${{\rm C}_{54}{\rm H}_{18}}$ & 1.07\footnotemark[3] & 1.35 &   -  &   -  \\
${{\rm C}_{54}{\rm H}_{19}}$ & 2.17\footnotemark[3] & 2.24 &   -  &   -  \\
\end{tabular}
\end{ruledtabular}
\footnotetext[1]{See, for example,
http://www.chemicool.com/elements/hydrogen.html.}
\footnotetext[2]{The electron affinities of coronene and coronene dimer
have been measured by Duncan {\em et al.} as reported in
Ref.(\onlinecite{Duncan}).}
\footnotetext[3]{This result corresponds to the small MIDI basis
type of calculation that is described in Ref.(\onlinecite{MIDI}).}
\label{ElectronAffinities}
\end{table}

Let us briefly discuss the results given in Table II.
The electron affinity of ${{\rm C}_{24}{\rm H}_{13}}$ cluster, i.e.,
the cluster with one hydrogen chemisorbed on top of one of the six
carbon atoms on the inner ring of coronene (see, Fig. 1) is 1.18 eV
larger than the electron affinity of coronene. Also, the value
for ${{\rm C}_{54}{\rm H}_{19}}$, that is, H on top of supercoronene
is 0.89 eV larger than the supercoronene electron affinity. Although,
only two cluster sizes have been studied, it seems that the difference is
approaching a limiting value close to the electron affinity of
free hydrogen, that is, close to 0.75 eV. 
If this were the case, the extra charge would be attracted by hydrogen
with a similar strength as in free space and the screening by the rest
of $\pi$-electrons of graphene would remain unnoticed. Nevertheless,
we show later that only part of the extra charge remains close to
the defect. Therefore, we assume that for an infinite system only a fraction
of 0.75 eV proportional to the localized charge would remain.

There is an alternative elaboration of the results given in Table I
focusing on the variation of the binding of a hydrogen atom on top
of a charged surface compared to the binding by the neutral one.
From this point of view, H binding energy increases from 0.59 eV to
1.77 eV on coronene and from 0.36 eV to 1.25 eV in supercoronene using
PC2 values in Table I\cite{binding}. This means that the charged systems
bind hydrogen about one eV stronger than the neutral ones.
From this perspective, the larger values of the electron
affinity obtained for hydrogenated clusters can be asigned
to a stronger hydrogen binding to graphene.

\begin{figure}
\includegraphics[width=\columnwidth]{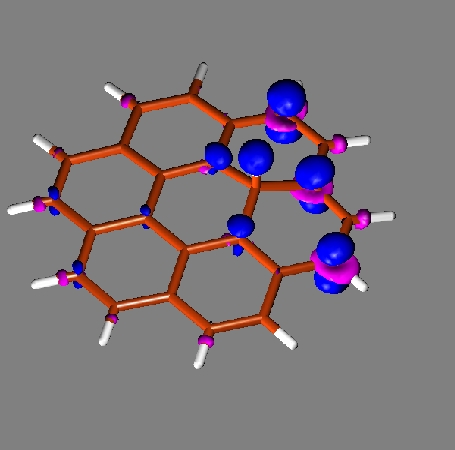}
\caption{(color online) Charge difference between the anion of coronene
and the neutral molecule. Density iso-contours of
$\pm$0.002 e/{\AA}$^3$ are represented.}
\label{resta}
\end{figure}

\begin{figure}
\includegraphics[width=\columnwidth]{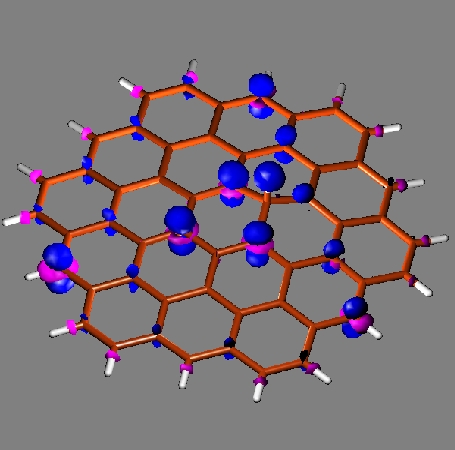}
\caption{(color online) Charge difference between the anion of supercoronene
and the neutral molecule. Density isocontours of
$\pm$0.002 e/{\AA}$^3$ are represented.}
\label{restaSC}
\end{figure}

\begin{figure}
\includegraphics[width=\columnwidth]{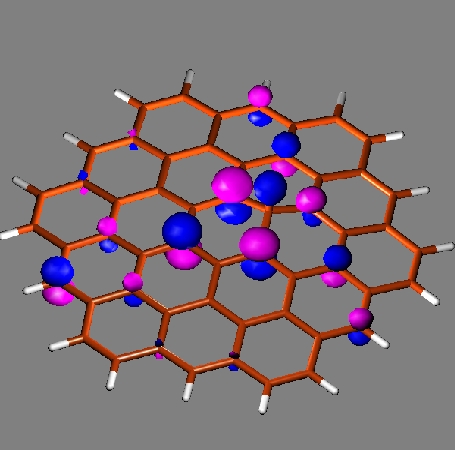}
\caption{(color online) HOMO orbital of supercoronene. Isocontours of
$\pm$0.05 {\AA}$^{-3}$ are represented.}
\label{homoSC}
\end{figure}

\begin{figure}
\includegraphics[width=\columnwidth]{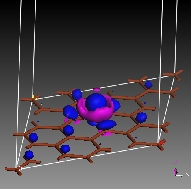}
\includegraphics[width=\columnwidth]{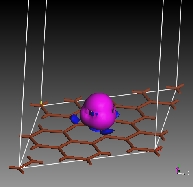}
\caption{(color online)
A $4 \times 4$ graphene supercell with a single adsorbed H atom
showing the bonding charge density first for the system with
one extra electron (upper panel) and, second, for the neutral
system (lower panel).
Density isocontours of $\pm$0.04 e/{\AA}$^3$ are given (blue and
violet respectively). The
accumulation of charge near H is seen to be the origin
for the cancelation of the extra spin. 
}
\label{fig4x4}
\end{figure}

The relative facility to move charge across the overall system implied
by the semi-metallic character of graphene and our 
results in Table II point towards the formation of a complex point defect
with an extra electron in the neighborhood of the chemisorbed hydrogen atom.
This picture is further supported by the spatial distribution of this
extra electron in the studied clusters as it is shown in 
Figs.~\ref{resta} and ~\ref{restaSC}.
Charge densities obtained with PC2 gaussian basis for coronene anion
and neutral coronene have been subtracted in Fig.~\ref{resta} using the
MOLDEN package\cite{MOLDEN}. The same difference for supercoronene
is given in Fig.~\ref{restaSC}. In both cases, the charge around "on top" H is
similar to the extra charge of the H anion. On the other hand, the spreading
positive and negative densities are similar but not equal in
coronene and supercoronene. A closer inspection reveals that
the extra electron is occupying the partially occupied HOMO level of the
corresponding neutral clusters. Fig.~\ref{homoSC} gives a picture of the HOMO
orbital of H on supercoronene that nicely explains the charge
difference previously shown in Fig.~\ref{restaSC}. 
It is interesting to notice, however, that although the extra charge clusters 
around the adsorbate, it is partly delocalized as it is expected from quantum 
mechanics principles. In fact, from independent tight-binding calculations in 
periodic systems we don't find a true {\it exponential} localization around 
the defect (see Appendix).
On the contrary, a percentage of the charge is extended all over 
the system (e.g., see Fig.~\ref{restaSC}). However, as our detailed 
quantum-mechanical calculations show, the net effect of the localized part 
is enough to quench the spin.

Our {\em ab initio} results on periodic extended systems fully support
the interpretation given in the previous paragraph.
We observe in Fig.~\ref{fig4x4} how the
extra charge is accumulated around the adsorbed H.
In this case, bonding charges for the neutral and charged supercells
are depicted in the upper and lower panels of the figure, respectively.
Bonding charges are defined as density charge differences between the
whole system and conveniently defined fragments. In our case, hydrogen
atom is one fragment while the graphene $4 \times 4$ supercell is the
second one. The upper difference integrates to one electronic charge
since the whole system is charged while the fragments are neutral.
We notice that a small part of the extra electron is on carbon atoms while
an important part attaches to hydrogen (the violet negative density
in the lower panel does not appears in the upper panel meaning a
positive contribution to the extra charge in the upper panel). 
The same picture is extracted from a Mulliken analysis of populations
around different atoms. In the neutral system, charge flows upon 
adsorption from hydrogen to graphene, so approximately $-0.63$ e is
located around hydrogen while the transferred charge resides
mostly around the closest carbon behind
hydrogen ($-0.33$ e). 
On the other hand, in the charged system we find $-1.42$ e around
hydrogen, while the carbon behind keeps nearly the same occupation
($-0.34$ e) and the rest of the charge is distributed over nearest
neighbors and next-nearest neighbors. Therefore, about 80\% of
the extra electron is located near the chemisorbed hydrogen. 
Along the same line, the H-C bond population analysis is about
three times larger for the anion, although the bond length is
nearly not affected.
Finally, integration of the spin density and the absolute value
of the spin density over the simulation cell give further 
support for this picture. In the neutral system these values
amount to $0.4$ and $0.5$ $\mu_{B}$ respectively, while in
the charged one decrease to values of 
amount to $4\times 10^{-6}$ and 
$5\times 10^{-4}$ $\mu_{B}$ respectively.
Therefore, the accumulation of charge around the adsorbed H and
the C nearest and next nearest neighbors plays the role to cancel
the extra spin polarization brought by the adsorption of H on the
clean graphene layer in accordance with the results obtained
on finite clusters.

There is a subtle chemical argument that helps the understanding of
our numerical results. In Ref.(\onlinecite{tryphenilmethyl}),
trivalent carbon atoms with an unpaired electron
were unraveled in the studied geometry of carbon tetrapod.
This carbon radicals were stabilized by steric protection giving
rise to unpaired localized electrons that
polarize the carbon neighborhood and explain the appearance of
magnetism in pure organic systems. The original paradigm
is tryphenilmethyl, synthesized by Gomberg in 1900\cite{Gomberg},
where a trivalent carbon atom is stabilized by three bonds to benzene
rings impeding the reaction with a similar molecule. Nevertheless,
the anion of tryphenilmethyl reacts with a proton
to form a strong C-H bond 
(heat of reaction amounts to 15.65 eV per molecule):
$$
{{\rm C}_{19}{\rm H}_{15}}^{-} + {\rm H}^{+} \rightarrow
{\rm C}_{19}{\rm H}_{16},
$$
producing a neutral non-magnetic molecule resembling the clusters that
we have studied here\cite{NIST}. Both the number of H atoms and electrons are
even allowing an easy shell closing and stabilization of the
resulting molecule. We can adapt the underlying chemistry of these
phenomena to the binding of H on graphene using the following argument:
it can be thought that
when H forms a covalent bond with a C atom of graphene the two
binding electrons are paired but additionally and due to the particular
topology of graphene lattice one $\pi$-electron becomes unpaired and,
therefore, spin polarized. Since there is no steric protection
around the defect (a kind of radical) any more or less free electron
of the system will flow into the defect to restore the
equilibrium between sublattices. Consequently, spin polarization
around chemisorbed hydrogen disappears. This qualitative
argument is fully supported by our total energy calculations
showing a gain in potential energy following the spin
neutralization (remember that the binding energy of H increases about
one eV for the charged system).

\section{Concluding Remarks}
From a detailed analysis based in first-principles DFT calculations 
we find that in the presence of an extra electron chemisorbed H plays 
to keep most of the extra charge in its vicinity. 
The electron affinities computed on finite cluster models seem to 
converge to a value that is somewhat smaller than the free atomic hydrogen
value of $0.75$ eV. Our calculations suggest that being
graphene a semimetal with zero density of states at the Fermi 
energy, screening of Coulomb interactions by the $\pi$-electrons
liquid is weak and allows the electron flow to 
sites where H is chemisorbed, forming a complex point defect. 
Accumulation of extra charge around the defect, otherwise
giving rise to spin polarization, works to quench it. 
We suggest that this physical effect is behind the difficulties to 
observe magnetism in graphene derived systems.

\begin{acknowledgments}
Financial support by the Spanish MICINN (MAT2006-03741,
MAT2008-1497, FIS2009-08744 and CSD2007-41) is gratefully acknowledged.
\end{acknowledgments}

\appendix*

\section{A tight-binding calculation}
A tight-binding model including carbon $\pi$-orbitals and hydrogen 1s orbital
allows a straightforward although approximate solution of the
topic covered in this paper. Since larger sizes can be studied,
qualitative conclusions on the scaling behavior of the charged system
can be obtained. This calculation fully supports our arguments based
on the computationally demanding {\it ab initio} study done in the
main part of the paper for smaller systems.
Although total energies are not attainable by the method, both the
density of states and the charge distribution of an extra electron
suggest the stability of a charged hydrogen atom bonded to graphene.

The p$_z$-p$_z$ hopping term of the Hamiltonian is -2.71 eV.
Comparing the ionization energy H (13.60 eV) to the
value for the methyl radical (9.84 eV) we take
the H level relative to the $\pi$ one as -3.76 eV. 
This is justified because of the similarity between graphene and 
methyl radical p$_z$ orbital. 
For the C-H coupling we have taken -5.39 eV based on a simple scaling argument.
A standard supercell procedure and a carefully BZ integration have
been used to get accurate density of states and local charge magnitudes.
Although very large supercells have been explored ($50 \times 50$),
we will give here results for a $10 \times 10$ system because it
shows almost converged results that are better depicted.

\begin{figure}
\includegraphics[width=\columnwidth]{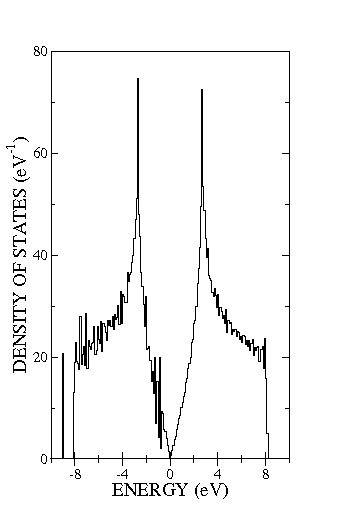}
\caption{Total density of states of a $10 \times 10$ supercell calculation
of a graphene layer containing one hydrogen atom forming a strong
covalent bond with one of the carbon atoms of the layer. Its integration
up the the Dirac point at E=0 gives one extra electron above half-filling.}
\label{dos}
\end{figure}

Fig.(\ref{dos}) shows the total density of states per spin of the
supercell. It integrates to 402 (total number of states of
the cell) and shows two salient features: 
(i) a deep level below the valence band describing
the bonding C-H orbital,  and (ii) strongly perturbed values below
Dirac's Point energy (0 in the clean system). Nevertheless, the most interesting
feature relative to this density of states (DOS)
is that it integrates to 202 up to the
Dirac energy (E=0), i.e., the complete occupation of the valence band describes
one extra electron above half-filling which is only 201 
(200 electrons corresponding
to the 200 $\pi$-orbitals of C atoms plus one electron contributed by H).
This means that when the Fermi level coincides with the Dirac point
of graphene, a region with a quimisorbed H atom becomes charged
by an extra electron. Since the full occupancy of the valence band
can be thought as closing electronic shells, some stabilization
of the system can be inferred from this property of the DOS.

\begin{figure}
\includegraphics[width=\columnwidth]{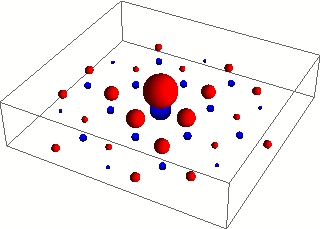}
\includegraphics[width=\columnwidth]{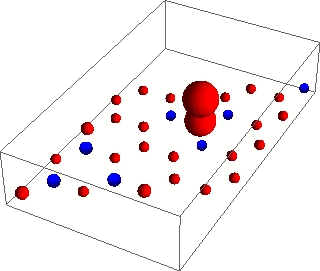}
\caption{(color online) Spatial distribution of the extra electron
charging the hydrogenated supercell. The sphere volume is proportional to
the extra number of electrons of the corresponding atom, red if positive and
blue if negative. The relevant part of the tight-binding results
obtained for a $10 \times 10$ supercell calculation is given in the
upper panel whereas {\em ab initio} results for the $4 \times 4$
supercell are shown in the lower panel. Last values are based on
Mulliken populations. Overall trends coincide but details differ:
both bonded H and C atoms get extra population when analyzed by the
Materials Studio (CASTEP) package but not for the semiempirical model, for example.
Although this representation of
the extra charge cannot be directly compared with isocountours
shown in Figs.(\ref{resta}) and (\ref{restaSC}), the overall
distribution looks similar.}
\label{mulliken}
\end{figure}

The ideas suggested by the DOS are confirmed by the spatial
distribution of the electronic charge induced by the addition of
an extra electron to the system. Fig.(\ref{mulliken}) (upper panel) has
been obtained subtracting the background charge of one electron per site
from a $10 \times 10$ supercell with an additional electron.
The resulting distribution can be compared with the corresponding
variations of Mulliken populations of our {\em ab initio} results
(lower panel). It shows that an important amount
of charge directly resides on the hydrogen atom although the underlying
carbon atom somewhat reduces the net value on the binding site
in the empirical model results.
Yet another important part of the electron lies close to this site
whereas the rest is spread over the whole supercell.
The most intriguing characteristic
of the charge distribution is its dual nature of localized around the
defect and extended over the whole system\cite{Dirac}.
In any case, a tight binding
model with a minimum number of orbitals reinforces our idea signaling
the tendency of electrons to remain in the neighborhood of bounded
hydrogen.

\newpage
\begin{thebibliography}{99}

\bibitem{Novoselov}
K.S. Novoselov, A.K. Geim, S.V. Morozov, D.Jiang, Y.Zhang, S.V. Dubonos,
I.V. Grigorieva, and A.A. Firsov,
Science {\bf 306}, 666 (2004).

\bibitem{Berger}
C. Berger, Z. Song, T. Li, X. Li, A.Y. Ogbazghi, R. Feng,
Z. Dai, A. N. Marchenkov, E.H. Conrad, P.N. First, and W.A. de Heer,
J. Phys. Chem. B {\bf 108}, 19912 (2004).

\bibitem{magnetism}
P. Esquinazi, A. Setzer, R. H\"ohne, C. Semmelhack, Y. Kopelevich,
D. Spemann, T. Butz, B. Kohlstrunk, and M. L\"osche,
Phys. Rev. B {\bf 66}, 024429 (2002);
K. Kusakabe and M. Maruyama, Phys.\ Rev.\ B {\bf 67}, 092406 (2003);
A.N. Andriotis, M. Menon, R.M. Sheetz, and L. Chernozatonskii,
Phys. Rev. Lett. {\bf 90}, 026801 (2003);
N. Park, M. Yoon, S. Berber, J. Ihm, E. Osawa, and D. Tomanek,
Phys. Rev. Lett. {\bf 91} 237204 (2003);
P.O. Lehtinen, A.S. Foster, Y. Ma, A.V. Krasheninnikov, and R.M. Nieminen,
Phys. Rev. Lett. {\bf 93}, 187202 (2004);
Y-W Son, M.L. Cohen and S.G. Louie, Nature {\bf 444}, 347 (2006).

\bibitem{HonGraphene}
Y. Miura, H. Kasai, W.A. Di\~no, H. Nakanishi and T. Sugimoto,
J. Phys. Soc. Jpn. {\bf 72}, 995 (2003).

\bibitem{stress}
P.L. de Andres and J.A. Verg\'es,
Appl. Phys. Lett. {\bf 93}, 171915 (2008).

\bibitem{HonGraphene2}
D.W. Boukhvalov, M.I. Katsnelson and A.I. Lichtenstein,
Phys. Rev. B {\bf 77}, 035427 (2008).

\bibitem{graphane}
D.C. Elias, R.R. Nair, T.M.G. Mohiuddin, S.V. Morozov, P. Blake, M.P. Halsall,
A.C. Ferrari, D.W. Boukhvalov, M.I. Katsnelson, A.K. Geim, and K.S. Novoselov,
Science {\bf 323}, 610 (2009).

\bibitem{lieb}
E.H. Lieb, Phys. Rev. Lett. {\bf 62}, 1201 (1989).

\bibitem{innombrables}
A summary of this explanation was recently published by
J. Fern\'andez-Rossier and J.J. Palacios,
Phys. Rev. Lett. {\bf 99}, 177204 (2007).

\bibitem{PAH}
J. A. Verg\'es, G. Chiappe, E. Louis, L. Pastor-Abia, and E. SanFabi\'an,
Phys. Rev. B {\bf 79}, 094403 (2009).

\bibitem{SW}
E.J. Duplock, M. Scheffler, and P.J.D. Lindan,
Phys. Rev. Lett. {\bf 92}, 225502 (2004).

\bibitem{GAMESS}
M.W. Schmidt, K.K. Baldridge, J.A. Boatz, S.T. Elbert, M.S. Gordon,
J.H. Jensen, S. Koseki, N. Matsunaga, K.A. Nguyen, S.J. Su, T.L. Windus,
M. Dupuis, J.A. Montgomery,
J. Comput. Chem. {\bf 14}, 1347 (1993).

\bibitem{MIDI}
J. Andzelm, M. Klobukowski, E. Radzio-Andzelm, Y. Sakai,
H. Tatewaki in {\em Gaussian basis sets for molecular calculations},
edited by S. Huzinaga (Elsevier, Amsterdam, 1984).

\bibitem{CCT}
T.H. Dunning, Jr., J. Chem. Phys. {\bf 90},  1007  (1989).

\bibitem{PCn}
F. Jensen,
J. Chem. Phys. {\bf 115}, 9113 (2001);
{\em ibid.} {\bf 116}, 7372 (2002).

\bibitem{B3LYP}
A.D. Becke, J. Chem. Phys. {\bf 98}, 5648 (1993);
C. Lee, W. Yang and R.G. Parr, Phys. Rev. B {\bf 37}, 785 (1988);
R. Colle and O. Salvetti, Theor. Chim. Acta {\bf 37}, 329 (1975).

\bibitem{Vanderbilt}
D. Vanderbilt, Phys. Rev. B, {\bf 41}, 7892 (1990).

\bibitem{monk}
H.J. Monkhorst and J.D. Pack, Phys. Rev. B, {\bf 13}, 5188 (1976).

\bibitem{Accelrys}
Materials Studio 4.4; http://www.accelrys.com.

\bibitem{Payne}
S.J. Clark, M. D. Segall, C.J. Pickard,
P.J. Hasnip, M. J. Probert, K. Refson and M. C. Payne,
Z. fuer Kristallographie {\bf 220}, 567 (2005).

\bibitem{Kohn}
W. Kohn and L. J. Sham, Phys. Rev., {\bf 140}, A1133 (1964).

\bibitem{EA}
Electron affinity is obtained as the opposite of the difference
between total energies of a molecule having an extra electron and the
neutral one. Table I always gives the charged molecule values in the
row following the corresponding neutral values.

\bibitem{Duncan}
M.A. Duncan, A.M. Knight, Y. Negishi, S. Nagao, Y. Nakamura,
A. Kato, A. Nakajima, and K. Kaya,
Chem. Phys. Lett. {\bf 309}, 49 (1999).

\bibitem{binding}
Using values given by the fifth column of Table I we get:
$27.21 \times (-922.1723-(-921.6516-0.4990))$ eV = -0.59 eV for coronene,
$27.21 \times (-922.2300-(-921.6660-0.4990))$ eV = -1.77 eV for negatively
charged coronene,
$27.21 \times (-2068.9095-(-2068.3971-0.4990))$ eV = -0.36 eV for
supercoronene, and
$27.21 \times (-2068.9919-(-2068.4469-0.4990))$ eV = -1.25 eV for negatively
charged supercoronene.

\bibitem{MOLDEN}
G. Schaftenaar and J.H. Noordik,
"Molden: a pre- and post-processing program for molecular and
electronic structures",
J. Comput.-Aided Mol. Design {\bf 14} 123 (2000).

\bibitem{tryphenilmethyl}
N. Park {\em et al.} in Ref.(\onlinecite{magnetism}).

\bibitem{Gomberg}
M. Gomberg, J. Am. Chem. Soc. {\bf 22}, 757 (1900).

\bibitem{NIST}
A better insight can be obtained from the figures shown in:
http://webbook.nist.gov/cgi/cbook.cgi?ID=B823\&Units=SI\&Mask=8.

\bibitem{Dirac}
The fact that the $\pi$-bands of graphene cross at only a pair of
{\bf k}-points at the Dirac energy is most probably the origin of this kind
of mixed soft localization.





\end{thebibliography}
\end{document}